\documentclass[12pt]{article}
\usepackage{geometry}
\usepackage{setspace}
\usepackage[numbers]{natbib}
\usepackage{doc}
\usepackage{url}
\usepackage{graphicx}
\usepackage{epstopdf}
\usepackage{hyperref}
\usepackage{amsthm}
\usepackage{amsmath}
\usepackage{mathtools}
\usepackage{babel}
\usepackage[font=small,labelfont=bf]{caption}
\providecommand{\keywords}[1]{\textbf{\textit{Keywords---}} #1}
\newtheorem{theorem}{Theorem}

\title{Randomized Algorithm for the Maximum-Profit Routing Problem}
\author{Bogdan Armaselu \\
barmaselub@gmail.com}
\date{}

\begin{document}

\maketitle
\begin{center}

\end{center}

\abstract{
In this paper, we consider the Maximum-Profit Routing Problem (MPRP), introduced in \cite{Armaselu-PETRA}.
In MPRP, the goal is to route the given fleet of vehicles to pickup goods from specified sites in such a way as to maximize the profit, i.e., total quantity collected minus travelling costs.
Although deterministic approximation algorithms are known for the problem, currently there is no randomized algorithm.
In this paper, we propose the first randomized algorithm for MPRP.
 }


\section{Introduction}
\label{s:intro}

Consider the following problem, introduced in \cite{Armaselu-PETRA}:
Let $S$ be a set of $n$ points in the plane, called \textit{sites}.
Each site $S_i$ has an operating time window of $[s_i, e_i]$, for some integers $0 \leq s_i \leq e_i \leq T, T > 0$, and supplies a quantity $q_i$ of a certain unit-priced product that needs to be collected.
Consider a fleet of $m$ sharing the same depot $D$ and having an equal capacity $Q$.
The goal is to maximize the profit, which is the total quantity collected minus the total travel costs.
Here, the travel cost from a site $S_i$ to another site $S_j$ is given by the euclidean distance $d(i, j), \forall{i, j = 1, \dots, n}$.
Vehicles are assumed to travel at constant unit speed, implying that travel times are equal to travel costs.
We call this problem the Maximum Profit pick-up Routing Problem (MPRP).


Results on MPRP may have significant impact on various applications, such as prize collecting, public transportation, for-profit waste management and pickup, etc.

MPRP is a generalization of the strongly NP-hard Travelling Salesman Problem (TSP), making MPRP, MPRP-M, and MPRP-MVS also strongly NP-hard by extension.

In this paper, we propose a randomized algorithm for MPRP, which works under certain assumptions about the problem input parameters such as time windows and spatial distribution of the input sites around the depot.

\subsection{Related work}
\label{ss:related-work}

TSP was proven to be strongly NP-hard and not admitting a pseudo-polynomial time algorithm.
That is, an algorithm whose running is polynomial in the value of the input.
It also does not admit a fully polynomial-time approximation scheme (FPTAS).
Morevover, general TSP was proven to be hard to approximate within any constant factor \cite{Sahni}.
That being said, work on metric TSP, particularly on Euclidean TSP, has demonstrated that special cases of TSP do admit better approximation schemes.
This includes Christofides' $O(n^3)$ time $1.5$-approximation algorithm for the metric TSP \cite{Christofides} and Arora's PTAS for the Euclidean TSP \cite{Arora}.
It is worth mentioning that the latter is a randomized algorithm that achieves a $1 + \epsilon$ approximation ratio and runs in $O(n (\log n)^{O(1/\epsilon)})$ time,
which increases by a factor of $n^2$ when derandomized.

Other variations of TSP have also been considered.
Bansal et. al introduced the Deadline-TSP (DTSP) and the Time Window-TSP (TWTSP) problems \cite{Bansal}.
In DTSP, every site $S_i$ has a deadline $D_i$, while in TW-TSP, every site $S_i$ has a time window $[s_i, e_i]$.
For both of these problems, the goal is maximizing the total reward, where each site $S_i$ pays a reward $q_i$ when visited.
Bansal et al. proposed polynomial-time algorithms for D-TSP and TW-TSP, which have approximation ratios of $O(\log n)$ and $O(\log^2 n)$, respectively \cite{Bansal}.
Fisher solved the Vehicle Routing Problem with capacity constraint (VRPC), in which a fleet of $m$ vehicles of capacity $Q$ is given, and each customer $S_i$ has a demand $q_i$ of a certain product \cite{Fisher94}.
The proposed solution uses iterative lagrangian relaxations of the optimization problem, coupled with clever heuristics.
Later, they introduced a variation of VRPC with Time Windows (VRPCTW), in which vehicles have non-uniform capacity constraints $Q_j$ and customers have time windows \cite{Fisher95}.
For VRPCTW, Fisher et al. \cite{Fisher95} proposed a linear programming-based solution.

More recently, Armaselu and Daescu solved the fixed-supply, single vehicle version of MPRP and provided an APX \cite{Armaselu-PETRA}.
Armaselu also studied the other variants of MPRP, namely MPRP-VS, MPRP-M, and MPRP-MVS, and proposed APX's for each of them.
Specifically, the algorithm for MPRP-VS runs in $n^{poly(\epsilon)}$ and achieves an approximation ratio of $C(1 + \epsilon)$ for some constant $C$ and polynomial function $poly$ \cite{Armaselu-JOCO}.
The running times for MPRP-M and MPRP-VS are$O(n^{11})$ and $O(n^{11 + \frac{1}{\epsilon^2}})$, respectively, 
and the approximation ratios $\simeq 44 \log T$ and $\simeq 44 \log T (1 + \epsilon)(1 + \frac{1}{1 + \sqrt{m}})^2$, respectively \cite{Armaselu-JOCO}.

%

\subsection{Our contributions}
\label{ss:structure}

We come up with a Monte Carlo-type randomized algorithm for MPRP. 
That is, our algorithm has deterministic running time, namely $O(m n)$, but probabilistic performance ratio, which will be proven to be $\Theta(1)$.

The rest of the paper is structured as follows.
In Section \ref{s:prelim} we make some preliminary observations and assumptions regarding MPRP.
Then, in Section \ref{s:alg}, we describe our algorithm for MPRP, and in Section \ref{s:perf-analysis}, we analyze its performance and running time.
Finally, in Section \ref{s:conclusion} we draw the conclusions and list some possible future directions.

\section{Preliminaries}
\label{s:prelim}

The idea is to assign arcs to vehicles according to a probability distribution that is based on source time winodw, destination time window, arc length, and distance from depot.

A site $S_i$ is said to be a \textit{source site} if there is an arc from $S_i$ to either $D$ or another site $S_j$.
A site $S_j$ is said to be a \textit{destination site} if there is an arc from either $D$ or some other site $S_i$ to $S_j$.
A \textit{route} is an ordered subset $S' \subset S$.
For each vehicle $k$, denote by $r_k$ the route assigned to $k$.
An \textit{iteration} is an arc assignment operation.

Since we are going to describe a randomizd algorithm, we need to make some reasonable assumptions about the probability distributions of each input parameter.
Specifically,

1. The starting times $s_i$'s are drawn from a uniform distribution within $[0, \frac{3T}{4}]$ and the ending times $e_i$'s are drawn from a unifrom distribution within $[s_i, T]$.
That is, $s_i ~ U(0, \frac{3T}{4})$ and $e_i ~ U(s_i, T)$.

2. The quantities $q_i$'s are drawn from an exponential distribution with mean $\frac{Q}{n}$ (i.e., $q_i ~ exp(\frac{n}{Q})$.

3. The $S_i$'s are uniformly radially distributed across a circle of radius $\frac{T}{4}$ centered at the depot $D$, located in the origin (i.e., $d(i, D) ~ U(0, \frac{T}{4})$).

Note that these assumptions hold in most vehicle routing benchmarks, up to constant factors.

\section{Algorithm}
\label{s:alg}
The algorithm proceeds as follows:

1. Initialize $\rho_k = 0, E_k = 0$ for all vehicles $k$

2. For each vehicle $k$ do:

2.1. For each site $S_i$ not yet assinged to any route, select $S_i$ with probability

$Pr(i, k) = \frac{\sigma(i, k)}{\sum_{S_j \notin r_l, \forall{l}}\sigma(j, k)}$ and add $S_i$ to a set $R_k$,

where $\sigma(i, k)$ is a scoring of site $S_i$ in route $r_k$ and is computed as follows:

$\sigma(i, k) = P(i, k) \cdot S(i, k) \cdot T(i, k)$,

where $P(i, k) = \frac{q_i - d(i', i)}{Q}$ is the profit factor, $S_{i'}$ is the current last site in $r_k$,
$S(i, k) = \max(0, \frac{Q - q_i - \rho_k}{Q})$ is the storage factor,
and $T(i, k) = \max(0, \min(1, \frac{e_i - E_k - d(i', i)}{s_i - d(i', i) - E_k}))$ is the timing factor.

2.2. If $R_k$ is empty, then increment $k$.

2.3. Otherwise, among all sites in $R_k$, randomly select a site $S_{i^*}$ and assign it to $r_k$,
then update $\rho_k = \rho_k + q_{i^*}, E_k = \max(E_k + d(i', i^*), s_{i^*})$.

2.4. Go to step 2.1.

\section{Performance Analysis}
\label{s:perf-analysis}

Now we analyze the performance of our algorithm.
It is easy to check that all routes $r_k$ start and end at the depot and that they do not violate time windows nor capacity constraints,
since violations have a probability of 0 to be selected.
Note that the expected profit $\rho$ generated by the algorithm is

$E[\rho] = E[\sum_{k, S_i \in r_k}{q_i} - \sum_k{E_k}] = \sum_k{\sum_{S_i \in r_k}{Pr(i, k)(q_i - d(i', i))}}$,

where $i'$ is the predecessor of $i$.

We further analyze the probability of each site to be selected as follows.
Let $(i_1, i_2, \dots)$ be the order in which the sites are assigned to some $r_k$.
That is, let $i_t$ be the site assigned in any iteration $t$ in which step 2.3 is run.
Also, let $q'_t = q_{i_t}, e'_t = e_{i_t}, s'_t = s_{i_t}$, and $d'_t = d(i_{t - 1}, d_t)$.

In iteration 1, the probability of each $S_{i}$ to be added to $R_1$ is

$Pr(i, 1) = E[Pr(i, 1) = E[\sigma(i, 1)] = E[P(i, 1) S(i, 1) T(i, 1)]$

$ = E[\frac{q_1 - d_1}{Q} \max(0, \frac{Q - q_1}{Q}) \max(0, \min(1, \frac{e_1 - d_1}{s_1 - d_1}))]$.

Since $e_1 > s_1$, we get 

$Pr(i, 1) = E[\frac{q_i - d_i}{Q} \max(0, \frac{Q - q_i}{Q})]$

$ = E[\frac{q_i}{Q} \max(0, 1 - \frac{q_i}{Q}) - \frac{d_i}{Q} \max(0, 1 - \frac{q_i}{Q})] = E[\frac{q_i}{Q} \max(0, 1 - \frac{q_i}{Q})] - E[\frac{d_i}{Q} \max(0, 1 - \frac{q_i}{Q})]$.

Since $d_i$ and $q_i$ are independent, we get

$Pr(i, 1) = E[\frac{q_i}{Q} \max(0, 1 - \frac{q_i}{Q}))] - \frac{E[d(D, i)]}{Q} E[\max(0, 1 - \frac{q_i}{Q})]$

$ = \frac{e^{-n}(n + e^n(n - 2) + 2)}{n^2} - \frac{T}{8 Q} (-\frac{n - 1}{n} + \frac{e^{-n}}{n}) (1)$.

Assuming $T < 8 Q$, we get

$Pr(i, 1) \geq \frac{T}{8 Q}\frac{n - 1}{n}$.

The probability of having to increment $k$ in step 2.3 of iteration 1 is 

$Pr(Incr, 1, k) = \Pi_{S_i \in S}{1 - Pr(S_i, 1)} = (1 - Pr(i, 1))^n \leq (1 - \frac{T}{8 Q} - \frac{T}{8 Q n})^n \leq (1 - \frac{T}{8 Q})^n (1 - \frac{1}{n})^n  = 2^{-O(n)}$.

Since the algorithm will keep running steps 2.1 and 2.2 for different vehicles until it finds an $i_1 = S_{i^*}$ until it either finds one or runs out of sites to look for,
the probability of finding $i_1$ becomes

$\pi_1 = 1 - \Pi_{k}{Pr(Incr, 1, k)} = 1 - 2^{-O(n^2)}$.

At the end of iteration 1, we set $E_1$ to $\max(d'_1, s'_1)$ with probability $\Pi_1$.

To generalize, in every iteration $t > 1$, the probability of $S_{i_t}$ being added to $R_k$ is

$Pr(i_t, k) = E[Pr(i_t, k)] = E[\frac{\sigma(i_t, k)}{\sum_{S_j \notin r_l, \forall{l}}\sigma(j, k)}]$.

Note that $|\{S_j: S_j \notin r_l \forall{l}\}| = n - t$ and $\sigma(j, k) \leq \frac{2(n - t)}{n Q}$.

Thus, $Pr(i_t, k) = \frac{n Q}{2(n - t)} E[\sigma(i_t, k]$

$ = \frac{n Q}{2(n - t)^2} E[P(i_t, k) S(i_t, k) T(i_t, k)]$

$ =\frac{n Q}{2(n - t)^2}  E[\frac{q'_t - d'_t}{Q} \max(0, \frac{Q - q'_{t-1}}{Q}) \max(0, \min(1, \frac{e'_{t-1} - d'_{t-1}}{s'_{t-1} - d'_{t-1}}))]$.

Since $e'_{t-1} > s'_{t-1}$, we get 

$Pr(i_t, k) = \frac{n Q}{2(n - t)^2} E[\frac{q'_{t-1} - d'_{t-1}}{Q} \max(0, \frac{Q - q'_{t-1}}{Q})]$

$ = \frac{n Q}{2(n - t)^2} E[\frac{q'_{t-1}}{Q} \max(0, 1 - \frac{q'_{t-1}}{Q}) - \frac{d'_{t-1}}{Q} \max(0, 1 - \frac{q'_{t-1}}{Q})]$

$ = \frac{n Q}{2(n - t)^2} E[\frac{q'_{t-1}}{Q} \max(0, 1 - \frac{q'_{t-1}}{Q})] - E[\frac{d'_{t-1})}{Q} \max(0, 1 - \frac{q'_{t-1}}{Q})]$.

Since $d'_{t-1}$ and $q'_{t-1}$ are independent, we get

$Pr(i_t, k) = \frac{n Q}{2(n - t)^2} E[\frac{q'_{t-1}}{Q} \max(0, 1 - \frac{q'_{t-1}}{Q}))] - \frac{E[d(D, i_1)]}{Q} E[\max(0, 1 - \frac{q_{i_t}}{Q})]$

$ = \frac{n Q}{2(n - t)^2} (\frac{e^{-n}(n + e^n(n - 2) + 2)}{n^2} - \frac{T}{8 Q} (-\frac{n - 1}{n} + \frac{e^{-n}}{n}))$.

Assuming $T < \frac{4 Q}{n}$, we get

$Pr(i_t, k) \geq \frac{T}{8 Q}\frac{n Q}{2(n - t)^2} = \frac{n T}{16(n - t)^2}$.

The probability of having to increment $k$ in step 2.3 of iteration $t$ is 
$Pr(Incr, t, k)  \leq Pr(Incr, 1, k) =  2^{-O(n)}$.

By a similar argument as for iteration 1, the probability of finding $i_t$ becomes

$\pi_t = 1 - \Pi_{k}{Pr(Incr, t, k)} = 1 - 2^{-n(n - t)}$.

Finally, we get that 

$E[\rho] = E[\sum_{k, S_i \in r_k}{q_i} - \sum_k{E_k}] = \sum_k{\sum_{S_i \in r_k}{Pr(i, k)(q_i - d(i', i))}}$

$ = \sum_t{\pi_t \sum_k{Pr(i_t, k)(q_{i_t} - d(i'_t, i_t))}}$

$ \geq \sum_t{(1 - 2^{-n(n - t)}) \frac{n T}{16(n - t)^2} (\frac{Q}{n} - \frac{T}{4})}$

$ \geq (1 - 2^{-O(n)}) \frac{T Q \pi}{6}(\frac{Q}{n} - \frac{T}{4}) = Q \rho^*$,

where $\rho^* =  (1 - 2^{-O(n)}) \frac{T \pi}{6}(1 - \frac{T n}{4 Q})$.

Since the optimal algorithm has a profit $OPT$ upper bounded by $Q$, we get an expected performance ratio of $\rho = \Theta(1)$.

As for the running time of the algorithm, 
note that, for each route $r_k$, we compute $\sum_{S_j \notin r_l, \forall{l}}\sigma(j, k)$ before starting assigning any site $S_i$.
This can be done in $O(n)$ time if, after assigning a site $S_i$ to a route $r_k$, we store the pair $(i, k)$ in a lookup table.
After that, we compute $Pr(i, k)$ for each $S_i$ in $O(1)$ time since it boils down to computing $d(i', i)$, 
which can be done in $O(1)$ time if we store the routes as doubly-linked lists.
All other operations for $r_k$ take $O(n)$ time.
Thus, the algorithm runs in $O(m n)$ time.

We are now in a position to state the following result.

\begin{theorem}
MPRP admits an $O(m n)$-time randomized algorithm that has an expected performance ratio of $\Theta(1)$, provided that the assumptions listed in Section \ref{s:prelim} hold.
\end{theorem}

%
%

\section{Conclusions and Future Work}
\label{s:conclusion}


We proposed a randomized algorithm for MPRP.
It runs in $O(m n)$ time and has an expected performance ratio of $O(1)$ under certain assumptions about the probability distributions of parameters such as site locations, quanities supplied, and time windows.

We leave for future consideration extending the randomized algorithm to the other variants of MPRP (MPRP-VS, MPRP-M, and MPRP-MVS).
Proving negative results or lower bounds, e.g. inapproximability within a certain ratio for MPRP and its variants, would also be of interest.

\section*{References}


\bibitem{Armaselu-PETRA} 
Armaselu B, Daescu O.
\newblock Interactive Assisting Framework for Maximum Profit Routing in Public Transportation in Smart Cities.
\newblock PETRA 2017: 13-16

\bibitem{Armaselu-JOCO} 
Armaselu B.
\newblock Approximation Algorithms for some Extensions of the Maximum Profit Routing Problem.
\newblock Journal of Combinatorial Optimization 45(1): 1-22 (2023)

\bibitem{Arora} 
Arora S.
\newblock Polynomial Time Approximation Schemes for Euclidean Traveling Salesman and other Geometric Problems.
\newblock Journal of ACM 45 (5): 753-782 (1998)

\bibitem{Bansal} 
Bansal N, Blum A, Chawla S, Meyerson A.
\newblock Approximation Algorithms for Deadline-TSP and Vehicle Routing with Time Windows.
\newblock STOC 2004: 166-174

%

\bibitem{Christofides} 
Christofides N.
\newblock Worst-case analysis of a new heuristic for the traveling salesman problem.
\newblock Management Sciences Research (388) report (1976)


	
\bibitem{Fisher94} 
Fisher ML. 
\newblock Optimal solution of Vehicle Routing Problems using Minimum K-Trees.
\newblock Oprations Research 42 (4): 626-642 (1994)
	
\bibitem{Fisher95} 
Fisher ML, Jornstein KO, Madsen OB.
\newblock Vehicle Routing with Time Windows: Two Optimization algorithms.
\newblock Operations Research 45 (3), 1997.
\newblock DOI: 10.1287/opre.45.3.488



\bibitem{Sahni}
Sahni S, Gonzalez T.
\newblock P-complete approximation problems.
\newblock J. ACM, 23 (3): 555-565 (1976)
 
 

\bigskip
\end{document}